# Quantifying Charge Extraction and Recombination Using the Rise and Decay of the Transient Photovoltage of Perovskite Solar Cells

*Lisa Krückemeier*[1,2], *Zhifa Liu*[1], *Thomas Kirchartz*[1,3*] *and Uwe Rau*[1,2*]

[1]IEK5-Photovoltaik, Forschungszentrum Jülich, 52425 Jülich, Germany
[2]Jülich Aachen Research Alliance, JARA-Energy and Faculty of Electrical Engineering and Information Technology, RWTH Aachen University, Schinkelstr. 2, 52062 Aachen, Germany
[3]Faculty of Engineering and CENIDE, University of Duisburg-Essen, Carl-Benz-Str. 199, 47057 Duisburg, Germany
E-mail: u.rau@fz-juelich.de
E-mail: t.kirchartz@fz-juelich.de

**Abstract**

The extraction of photogenerated charge carriers and the generation of a photovoltage belong to the fundamental functionalities of any solar cell. These processes happen not instantaneously but rather come with finite time constants, e.g., a time constant related to the rise of the externally measured open circuit voltage following a short light pulse. The present paper provides a new method to analyze transient photovoltage measurements at different bias light intensities combining rise and decay times of the photovoltage. The approach uses a linearized version of a system of two coupled differential equations that is solved analytically be determining the eigenvalues of a 2 × 2 matrix. By comparison between the eigenvalues and the measured rise and decay times during a transient photovoltage measurement, we determine the rates of carrier recombination and extraction as a function of bias voltage and establish a simple link between their ratio and the efficiency losses in the perovskite solar cell.

## I. Introduction

The efficiency of halide perovskite solar cells has been continuously rising over the past decade to values above 25%[1-6] that are now approaching the efficiencies of crystalline Si solar cells.[7-9] Future technological development will have to deal with issues of device stability[10, 11] but also thrive to further minimize efficiency-limiting loss processes in the bulk and at interfaces within the cell stack.[12-14] The identification and understanding of electrical losses will require to a large degree the ability to characterize solar cells and multilayer stacks with a variety of steady-state,[12, 15-18] time-domain[19-27] and frequency-domain techniques[28-32] that are sensitive to the transport and recombination of charge carriers. Especially, time- and frequency-domain techniques offer a large amount of information on dynamic processes in the solar cell,[19, 30, 33] while posing a substantial challenge in terms of the complexity of data analysis.[19] In case of the often lowly doped or intrinsic halide perovskites,[34] the kinetics of the charge carrier decay in response to a laser pulse in a typical time-domain measurement shows a variety of different features that would have to be numerically simulated in order to accurately extract information from the raw data.[19, 35, 36] As this can only be done by few and often costly simulation programs and requires time-consuming fitting procedures, a thorough analysis is rarely performed in literature.[37] Thus, work that is primarily aimed at device and process optimization usually analyzes the transients using mono-exponential or bi-exponential fits, while only a minority of method-focused publications delves deeper into the physics of understanding the decays.[26, 27, 33, 38-40]

Here, we show how to use relatively simple analytical solutions to systems of differential equations to extract the key performance-limiting parameters in halide perovskite solar cells from the rise and decay of the transient photovoltage in response to a laser pulse. The results combine the simplicity and comprehensibility of analytical equations with the multiplicity of physical phenomena that occur during a transient experiment on a perovskite solar cell. Transient experiments contain information on phenomena such as recombination and extraction of charge carriers that are difficult to disentangle from each other using traditional approaches.[41] Furthermore, capacitive charging and discharging of electrodes affects the rise and decay times of transients thereby adding complexity to the data analysis.[26, 42] We solve this problem by creating a numerical model, linearizing it around a bias condition and then solving it analytically by determining the eigenvalues of a $2 \times 2$ matrix. The model yields two time constants (the inverse eigenvalues), one for the rise and one for the decay of the photovoltage after the pulse. These two time constants can be experimentally determined as a function of light intensity. The application of the model to experimental data allows us to derive a time constant for recombination and one for charge extraction, whereby the ratio of these two time constants is directly related to the solar cell efficiency. The results provide a significant progress relative to previous approaches[25-27, 38, 41, 42] for the analysis of transient photovoltage data as we explicitly include the rise time[43] as a source of information, consider a voltage dependent resistance of the charge-transport layers and provide a comprehensible analytical solution that can be conveniently compared with experimental data. We note further that the two-component model introduced here is generic for the rise and the decay of the photovoltage in any solar cell.

## II. Experimental Data

Figure 1 shows experimental data from small-signal transient photoluminescence and transient photovoltage measurements, being recorded at one integrated setup illustrated as previously discussed in ref.[42]. The internal voltage $V_{int}$ results from small-signal TPL measurements and the corresponding external voltage decays $V_{ext}$ from TPV at different bias light intensities. For TPL the detection is done with a gated-CCD camera as described in the method section in the Supplementary Information. The internal- and external-voltage curves at four different bias levels are plotted in Figure 1c-j. The graphs on the left-hand side show the internal and external voltage curves on a logarithmic time scale, whereas the results on the left-hand side are plotted on a linear time axis. This dataset belongs to a solution-processed $CH_3NH_3PbI_3$ (MAPI) solar cell, with the layer stack glass/ITO/PTAA/MAPI/PCBM/BCP/Ag, whereby PTAA is poly(triarylamine), PCBM is phenyl-$C_{61}$-butyric acid methyl ester, and BCP is bathocuproine. The general properties and performance of these devices were previously discussed in ref.[19, 42, 44, 45]. In this example, the external voltage needs around several hundred nanoseconds to build up and reach its maximum voltage. Furthermore, the rise time shifts to longer times for lower steady state excitation conditions. Note that also the decays of the internal and external voltage do not necessarily need to overlap, as the additional example of Supplementary Figure 1 shows. In conclusion, these experimental results demonstrate that the exchange of the charge carriers happens with a finite speed and is possibly slow enough to interfere with the classical lifetime analysis of the TPV data. Furthermore, it becomes apparent that the internal excess voltage $\Delta V_{int}$ can be several times larger than the external excess voltage $\Delta V_{ext}$.

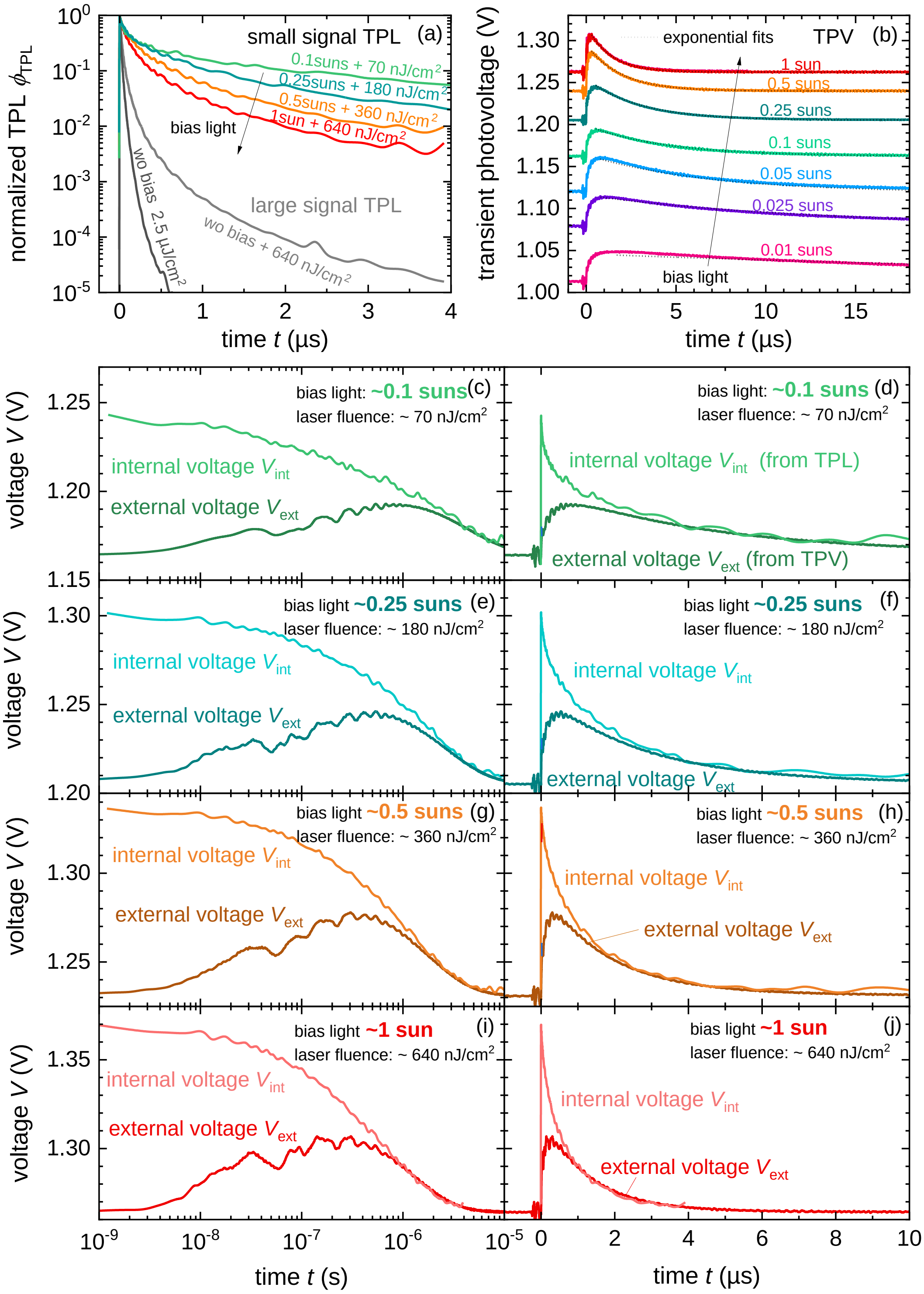

**Figure 1**: Experimental data of (a) large- and small-signal transient PL measurements and (b) transient photovoltage decays at different bias light intensities, measured on a MAPI solar cell. (c)-(j) Experimental data of the internal voltage $V_{\mathrm{int}}$ resulting for small-signal TPL measurements and the decays of external voltage $V_{\mathrm{ext}}$ from TPV at different bias light intensities, measured on a MAPI solar cell. The panels on the right show the data on a logarithmic times scale, whereas the panels on the left have a linar times scale.

## III. Analysis of the Two-Component Model

The theoretical description of the rise and decay of the externally measured photovoltage of a solar cell requires a minimum of two variables, one for the concentration of photogenerated charge carriers inside the absorber of the solar cell and one for the externally measured photovoltage. These two components are described with the help of a system of two coupled first order differential equations. These two equations are also required mathematically to describe a system where the time derivative changes its sign, i.e., a system with a rise followed by a decay.

The first differential equation describing the kinetics of charge carrier concentration $n$ in the absorber contains the terms for recombination as well as a term for the exchange current flowing to or from the electrodes. If this current is negative, it will reduce the carrier density in the absorber and increase the charge density on the electrodes. Alternatively, if the current is positive, the electrodes will inject charge carriers back into the absorber layer. For a perfectly symmetric system in high-level injection, i.e. if the electron concentration $n$ equals the hole concentration $p$ and the external voltage $V_{\mathrm{ext}}$ drops to equal amounts over the electron and over the hole contact, the first equation reads

$$\begin{aligned}\frac{dn}{dt} &= -k_{\mathrm{rad}} n^2 - \frac{n}{\tau_{\mathrm{SRH}}^{\mathrm{eff}}} - \frac{J_{\mathrm{exc}}}{qd} + G \\ &= -k_{\mathrm{rad}} n^2 - \frac{n}{\tau_{\mathrm{SRH}}^{\mathrm{eff}}} - \frac{S_{\mathrm{exc}}}{d}\left(n - n_i \exp\left(\frac{qV_{\mathrm{ext}}}{2kT}\right)\right) + G\end{aligned} \quad (1)$$

where $k_{\mathrm{rad}}$ is the radiative recombination coefficient, $\tau_{\mathrm{SRH}}^{\mathrm{eff}}$ is the effective Shockley-Read-Hall (SRH) recombination lifetime, $q$ the elementary charge, $d$ the thickness of the absorber, and $G$ is the generation rate. The quantity $J_{\mathrm{exc}}$ denotes the exchange current density and $S_{\mathrm{exc}}$ is an exchange velocity in units of cm/s that defines how quickly electrons are extracted from the absorber through the electron transport layer (or holes through the hole transport layer) and subsequently charge up the respective electrode. The ratio $d/S_{\mathrm{exc}}$ between the thickness $d$ and the exchange velocity $S_{\mathrm{exc}}$ defines a time constant for the increase and decrease of the carrier densities (per unit volume, assumed constant over a thickness $d$) caused by the exchange current flowing to or from the electrodes. Note that the way how the exchange current in equation (1) is linked to the carrier concentration and the external voltage is derived from the principle of detailed balance[46] and is also compatible with an extended equivalent circuit model for solar cells.[47] Equation (1) may also be rewritten as the change of the quasi-Fermi-level splitting $\Delta E_{\mathrm{F}}$ as

$$n(t) = p(t) = n_{\mathrm{i}} \exp\left(\frac{\Delta E_F(t)}{2k_{\mathrm{B}}T}\right) = n_{\mathrm{i}} \exp\left(\frac{qV_{\mathrm{int}}(t)}{2k_{\mathrm{B}}T}\right) \tag{2}$$

where we have replaced $\Delta E_{\mathrm{F}}$ by $qV_{\mathrm{int}}$, bearing in mind that this internal voltage $V_{\mathrm{int}}$ does not represent a voltage in the electrostatic sense but rather is used to conveniently relate the chemical potential of charge carriers within the device to the true, externally measurable, electrostatic potential given by $V_{\mathrm{ext}}$. With $dV_{\mathrm{int}}/dn = 2kT/(qn)$ the relation between the time derivative of $V_{\mathrm{int}}$ and the carrier concentration $n$ reads

$$\frac{dV_{\mathrm{int}}}{dt} = \frac{2kT}{qn}\left[-k_{\mathrm{rad}}n^2 - \frac{n}{\tau_{\mathrm{SRH}}^{\mathrm{eff}}} - \frac{S_{\mathrm{exc}}}{d}\left(n - n_i \exp\left(\frac{qV_{\mathrm{ext}}}{2kT}\right)\right) + G\right]. \tag{3}$$

The charge density $\sigma_{\mathrm{n}}$ per area on the n-type electrode will change according to the flux of electrons from/to the absorber described by the third term on the right-hand side of equation (1). We therefore get

$$\frac{d\sigma_n}{dt} = qS_{\mathrm{exc}}\left(n - n_{\mathrm{i}} \exp\left(\frac{qV_{\mathrm{ext}}}{2kT}\right)\right). \tag{4}$$

Here, the division by $d$ is missing because the equation describes a charge density per unit area (not a carrier concentration per unit volume as in equation (5)).

Eventually, we want to know how the external voltage is changing. We know that any change in external voltage requires a change in charge density per area according to $V_{\mathrm{ext}} = \sigma_n / C_{\mathrm{area}}$ where $C_{\mathrm{area}}$ denotes the capacitance of the junction. Thus, the time derivative of the external voltage follows

$$\frac{dV_{\mathrm{ext}}}{dt} = \frac{qS_{\mathrm{exc}}}{C_{\mathrm{area}}}\left(n - n_{\mathrm{i}} \exp\left(\frac{qV_{\mathrm{ext}}}{2kT}\right)\right). \tag{5}$$

With this, equations (1) and (4) form a system of two coupled non-linear differential equations in a picture that uses the carrier concentration and the surface charge density as variables. Likewise, we may use equation (2) to express the exchange current density as a function of external and internal voltage via

$$J_{exc} = qS_{exc}\left[n - n_{\mathrm{i}} \exp\left(\frac{qV_{\mathrm{ext}}}{2kT}\right)\right] = qS_{exc}n_{\mathrm{i}}\left[\exp\left(\frac{qV_{\mathrm{int}}}{2kT}\right) - \exp\left(\frac{qV_{\mathrm{ext}}}{2kT}\right)\right]. \tag{6}$$

With this, equations (3) and (5) represent an alternative system of differential equations using the voltages as variables. Furthermore, equation (6) shows that the exchange between absorber and electrodes is a net flux of electrons towards the electrode if $\Delta E_{\mathrm{F}}/q = V_{\mathrm{int}} > V_{\mathrm{ext}}$ while we have a net flux towards the absorber once $V_{\mathrm{int}} < V_{\mathrm{ext}}$. As long as $S_{\mathrm{exc}}$ is fairly high, which corresponds to a high mobility or conductivity of the electron and hole transport layers, the difference between $V_{int}$ and $V_{ext}$ will be small for a given current. In contrast, for small values of $S_{exc}$ the difference of $\Delta E_F$ and $qV_{ext}$ will be rather large. If the system is in a steady-state open-circuit situation where no current is flowing, we have $V_{int} = V_{ext}$.

The model described so far has a clear advantage relative to our recent approach because it explicitly considers the charging of the electrodes and the differences between external voltage and internal quasi-Fermi level splitting, expressed as internal voltage. Therefore, the model can be used to simulate e.g. the rise of the photovoltage and not only the decay. However, it has the disadvantage that we do not (yet) have an explicit analytical solution for the resulting time constants anymore. This is due to the fact that the model is based on a system of two coupled, non-linear differential equations. However, as long as we are considering a small-signal method as is typically the case for transient photovoltage measurements, it is possible to linearize the model shown above around a certain bias voltage $V_{bias}$ such that we go from a system of non-linear differential equations to a system of two linear differential equations. Such a system has analytical solutions implying that we can combine the simplicity of an analytical solution with the explicit consideration of a finite rate of charge extraction. The linearized version of equations (3) and (5) can be written as a matrix equation as

$$\frac{d}{dt}\begin{pmatrix}\delta V_{int}\\ \delta V_{ext}\end{pmatrix} = \begin{pmatrix} -\left(\frac{1}{\tau_{SRH}^{eff}} + 2k_{rad}n_{bias} + \frac{S_{exc}}{d}\right) & \frac{S_{exc}}{d} \\ \frac{S_{exc}n_{bias}}{dn_Q} & -\frac{S_{exc}n_{bias}}{dn_Q}\end{pmatrix}\begin{pmatrix}\delta V_{int}\\ \delta V_{ext}\end{pmatrix}. \tag{7}$$

Here, $n_{bias}$ is the carrier density at the bias condition of the small signal measurement, i.e. the concentration just before the laser pulse. Given that TPV is measured at open circuit, this carrier density follows $n_{bias} = p_{bias} = n_i \exp(qV_{int}/2k_BT) = n_i \exp(qV_{ext}/2k_BT)$ within the logic of our model.

This matrix equation can be solved and yields for the additional external voltage (relative to the bias voltage $V_{bias}$)

$$\delta V_{ext}(t) = V_{ext}(t) - V_{bias} = \delta V_0\left(\exp\left(-\frac{t}{\tau_{decay}}\right) - \exp\left(-\frac{t}{\tau_{rise}}\right)\right), \tag{8}$$

where the rise and decay times are given by

$$\tau_{rise} = \frac{2}{k_1 + k_2 + k_3 + \sqrt{(k_1 + k_2 + k_3)^2 - 4k_1k_3}} \tag{9}$$

and

$$\tau_{\text{decay}} = \frac{2}{k_1 + k_2 + k_3 - \sqrt{(k_1 + k_2 + k_3)^2 - 4k_1 k_3}},$$

(10)

whereby $k_1 = 1/\tau_{\text{SRH}}^{\text{eff}} + 2k_{\text{rad}} n_{\text{bias}}$, $k_2 = S_{\text{exc}}/d$, and $k_3 = S_{\text{exc}} n_{\text{bias}}/(d n_{\text{Q}})$. Note that the prefactor $\delta V_0$ in equation (8) is an arbitrary but small prefactor with unit volts that must be small enough to ensure that the linearization used to obtain equation (8) is justified.

The three inverse time constants $k_1$, $k_2$ and $k_3$ originate from the entries of the matrix in equation (7), whereby $k_1$ represents recombination. The other two can be understood as inverse RC constants, where $k_2$ is given by the resistance of charge extraction via the charge transfer layers multiplied with the chemical capacitance[31, 48] of the perovskite layer and the inverse time constant $k_3$ is given by the RC formed by the resistance of charge extraction multiplied with the electrode capacitance of the solar cell. Solving the matrix equation (7) also yields an expression for the additional internal voltage (relative to the bias voltage $V_{\text{bias}}$)

$$\delta V_{\text{int}}(t) = V_{\text{int}}(t) - V_{\text{bias}} = \delta V_0 \left( h_1 \exp\left( -\frac{t}{\tau_{\text{decay}}} \right) + h_2 \exp\left( -\frac{t}{\tau_{\text{rise}}} \right) \right), \quad (11)$$

with the two additional prefactors

$$h_1 = -\frac{k_1 + k_2 - k_3 - \sqrt{(k_1 + k_2 + k_3)^2 - 4k_1 k_3}}{2k_3} \quad (12)$$

and

$$h_2 = \frac{k_1 + k_2 - k_3 + \sqrt{(k_1 + k_2 + k_3)^2 - 4k_1 k_3}}{2k_3}. \quad (13)$$

## IV. Results from the Analytical Two-Component Model

Equations (8) and (11) allow us to calculate analytical solutions of the additional internal voltage $\delta V_{\text{int}}(t)$ and external voltage $\delta V_{\text{ext}}(t)$. The former would be the equivalent of a small signal transient photoluminescence curve, while the latter corresponds to a transient photovoltage signal. Examples of the respective analytical solutions of the normalized excess voltages $\delta V_{\text{int}}(t)/\delta V_0$ are shown in the upper two panels of Figure 2 and 3. With the help of these two Figures, we will successively discuss the influence of two important parameters on the $\delta V(t)/\delta V_0$ curves, namely the exchange velocity $S_{\text{exc}}$ and the electrode capacitance per area $C_{\text{area}}$.

Figures 2 (a) and (b) visualize the analytical solutions of the matrix equation for a variation of the exchange velocity $S_{\text{exc}}$ at two different bias levels of 1.1 V and 1.25 V. The normalized external voltage $\delta V_{ext}(t)/\delta V_0$ curves are plotted as solid lines, whereas the corresponding normalized internal voltage decays $\delta V_{int}(t)/\delta V_0$, being calculated via equation (11), are depicted as dotted lines. For the calculation of these graphs, we used three different exchange velocities $S_{exc}$, a high value of $10^3$ cm/s (blue), a medium value of 100 cm/s (red) and a small one of $S_{exc}$ = 10 cm/s (grey). The exchange velocity $S_{exc}$ depends on the properties of the charge-transport layers, such as the mobility or electric field distribution. Accordingly, the value of $S_{exc}$ determines how quickly electrons or holes are extracted or injected through the two transport layers. The other additional parameters, being necessary to calculate the voltage curves are listed in the figure caption of Figure 2.

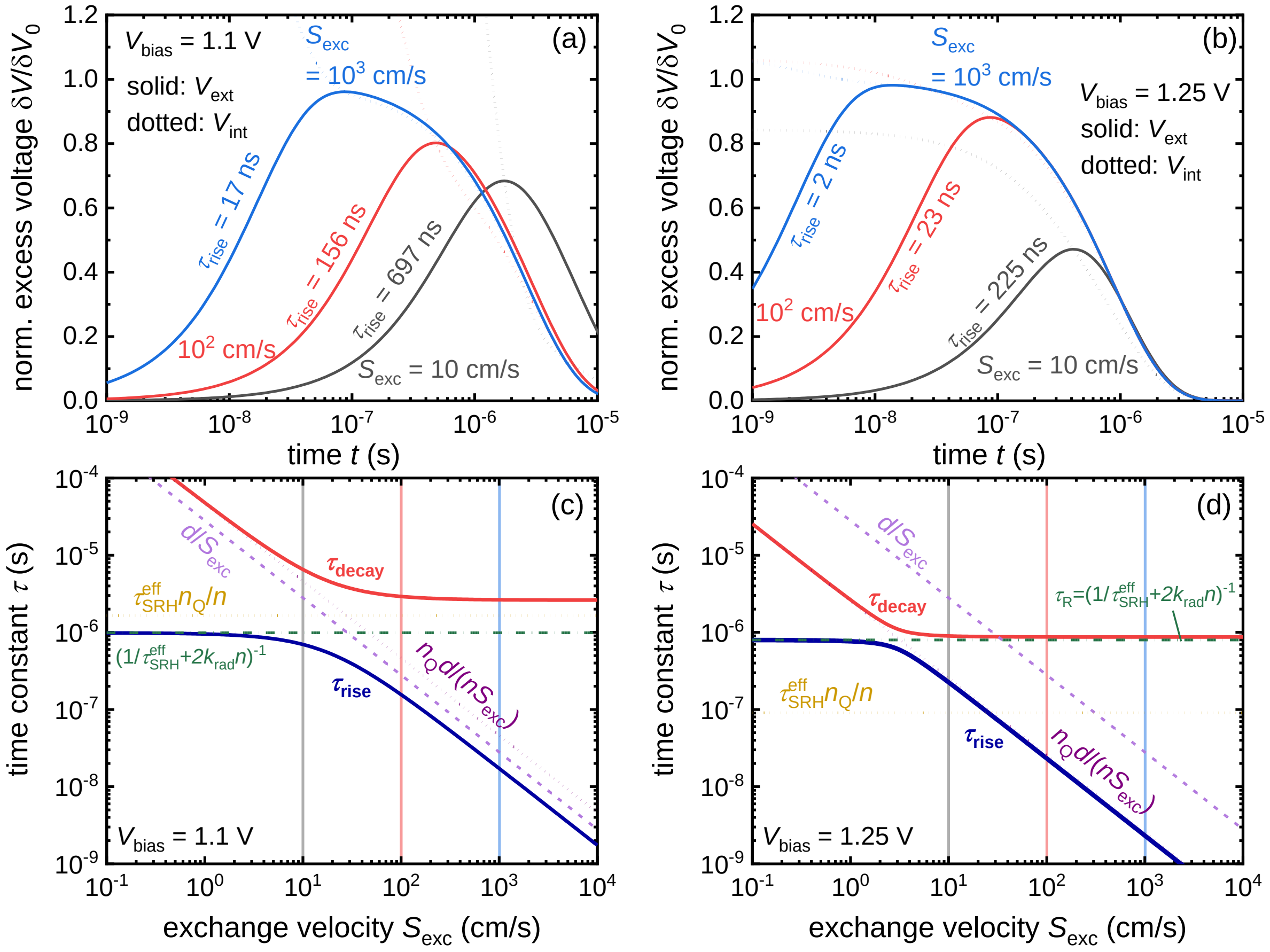

**Figure 2**: (a, b) Analytical solutions of the normalized internal excess-voltage decay $\delta V_{int}(t)/\delta V_0$ (dotted lines) and the respective normalized external voltage $\delta V_{ext}(t)/\delta V_0$ curve (solid lines) at (a) 1.1 V and (b) 1.25 V bias voltage, calculated for three exchange velocities $S_{exc}$, namely 1000 cm/s (blue), 100 cm/s (red) and 10 cm/s (grey). The rise of the normalized external voltage is prolonged for smaller exchange velocities. Additional parameters used for the calculations are a SRH lifetime $\tau^{eff}_{SRH} = 1\ \mu s$, a radiative recombination coefficient $k_{rad}$ of $5\times10^{-11}$ cm$^3$/s, an absorber-layer thickness $d$ of 280 nm, an intrinsic carrier concentration $n_i$ of $8.05\times10^4$ cm$^{-3}$ and a capacitance per area $C_{area}$ = 20 nF/cm$^2$. (c,d) Development of the decay time $\tau_{decay}$ and rise time $\tau_{rise}$ of the external voltage, defined by equation (9) as a function of the exchange velocity $S_{exc}$. Furthermore, additional time constants related to the eigenvalues of the matrix are depicted as guide to the eye.

Our first observation is that the analytical solution of the two-component model reflects the basic trend of the internal and external voltage curves from experiment. The normalized internal excess voltage $\delta V_{\text{int}}(t)/\delta V_0$ is highest at the very beginning and continues to decrease over time until it reaches the initial bias voltage. In contrast, the response of the normalized external excess voltage $\delta V_{\text{ext}}(t)/\delta V_0$ to the small-signal excitation is delayed. It increases slowly over time, reaches its maximum and only then begins to decay. Thus, this model is able to reproduce the general features of the TPV transient. Furthermore, the variation of the exchange velocity $S_{\text{exc}}$, shows that rise of the additional external voltage is prolonged for smaller exchange velocities and the maximum of the curves shifts to longer times. This trend occurs at both shown bias levels. The associated rise times $\tau_{\text{rise}}$, calculated via equation (9), are indicated next to the curves and quantify the described trend. The decay of $\delta V_{\text{ext}}(t)/\delta V_0$, on the other hand, is affected differently by the variation of exchange velocity for the two bias levels. In panel 2(a) at the lower injection level of 1.1 V, a change in $S_{\text{exc}}$ also affects the decay of the curves, whereas at 1.25 V in panel 2(b) all three decays of the normalized external voltage $\delta V_{\text{ext}}(t)/\delta V_0$ overlap. Note that the decay time $\tau_{\text{decay}}$, as well as the rise time $\tau_{\text{rise}}$ are injection-level dependent and therefore differ between Figure 2(a) and (b). At the higher bias voltage, the respective rise time $\tau_{\text{rise}}$ is shorter and increases by an order of magnitude when the exchange velocity decreases by an order of magnitude, thus follows a power law function while this inverse proportionality disappears in case of the lower bias voltage. To better understand the behaviour, Figures 2(c) and (d) directly depict the rise time $\tau_{\text{rise}}$ and decay time $\tau_{\text{decay}}$ as a function of the exchange velocity $S_{\text{exc}}$ for the two different bias levels. Furthermore, additional time constants related to the eigenvalues of the matrix and the inverse coefficients $1/k_1$, $1/k_2$, and $1/k_3$ are shown as guide to the eye. In addition, the parameter sets of the three different examples from the top figures are marked as well by thin, vertical lines.

The rise-time curve $\tau_{\text{rise}}$, shown as thick, blue line, can be divided into two sections. For low exchange velocities $S_{\text{exc}}$ the rise time is constant and determined by the recombination lifetime, namely $\tau_{\text{R}} = \left(1/\tau_{\text{SRH}}^{\text{eff}} + 2k_{\text{rad}}n_{\text{bias}}\right)^{-1}$ (green, dashed line). If the exchange is fast in comparison to recombination the rise time gets shorter with increasing exchange velocity $S_{\text{exc}}$. At the lower injection level of 1.1 V and high exchange velocities, the rise time depends on the ratio of $d/S_{\text{exc}}$ (lilac, dotted line). Thus, in this case $\tau_{\text{rise}}$ is dominated by the resistance of charge extraction through the charge-transfer layers and the chemical capacitance of the perovskite layer, which represent a $RC$-time constant. At high bias voltages (in our example: 1.25 V) as shown in Figure 2(d), the rise time $\tau_{\text{rise}}$ follows the $RC$-time constant $dn_{\text{Q}}/\left(S_{\text{exc}}n_{\text{bias}}\right)$ which is formed by the resistance of charge extraction multiplied with the electrode capacitance of the solar cell. The decay time $\tau_{\text{decay}}$, being calculated by using equation (10), is shown as thick, red line. In the regime where charge transfer is fast in comparison with recombination, the decay time of the external voltage is constant and does not depend on $S_{\text{exc}}$. Depending on the bias level the corresponding saturation value of $\tau_{\text{decay}}$ is set either by the time constant $\tau_{\text{SRH}}^{\text{eff}} n_{\text{Q}} / n_{\text{bias}}$ of the electrode being discharged via recombination in the perovskite or - at high bias levels - by recombination $\tau_{\text{R}} = \left(1/\tau_{\text{SRH}}^{\text{eff}} + 2k_{\text{rad}} n_{\text{bias}}\right)^{-1}$. If the exchange velocity is small, charge transfer also effects the decay time $\tau_{\text{decay}}$ of the normalized external voltage $\delta V_{\text{ext}}(t)/\delta V_0$, which then approaches $dn_{\text{Q}} / \left(S_{\text{exc}} n_{\text{bias}}\right)$ (purple, dashed line). This discussion shows that depending on the injection level and parameter set, different time constants determine both the rise and fall of the small-signal TPV curve.

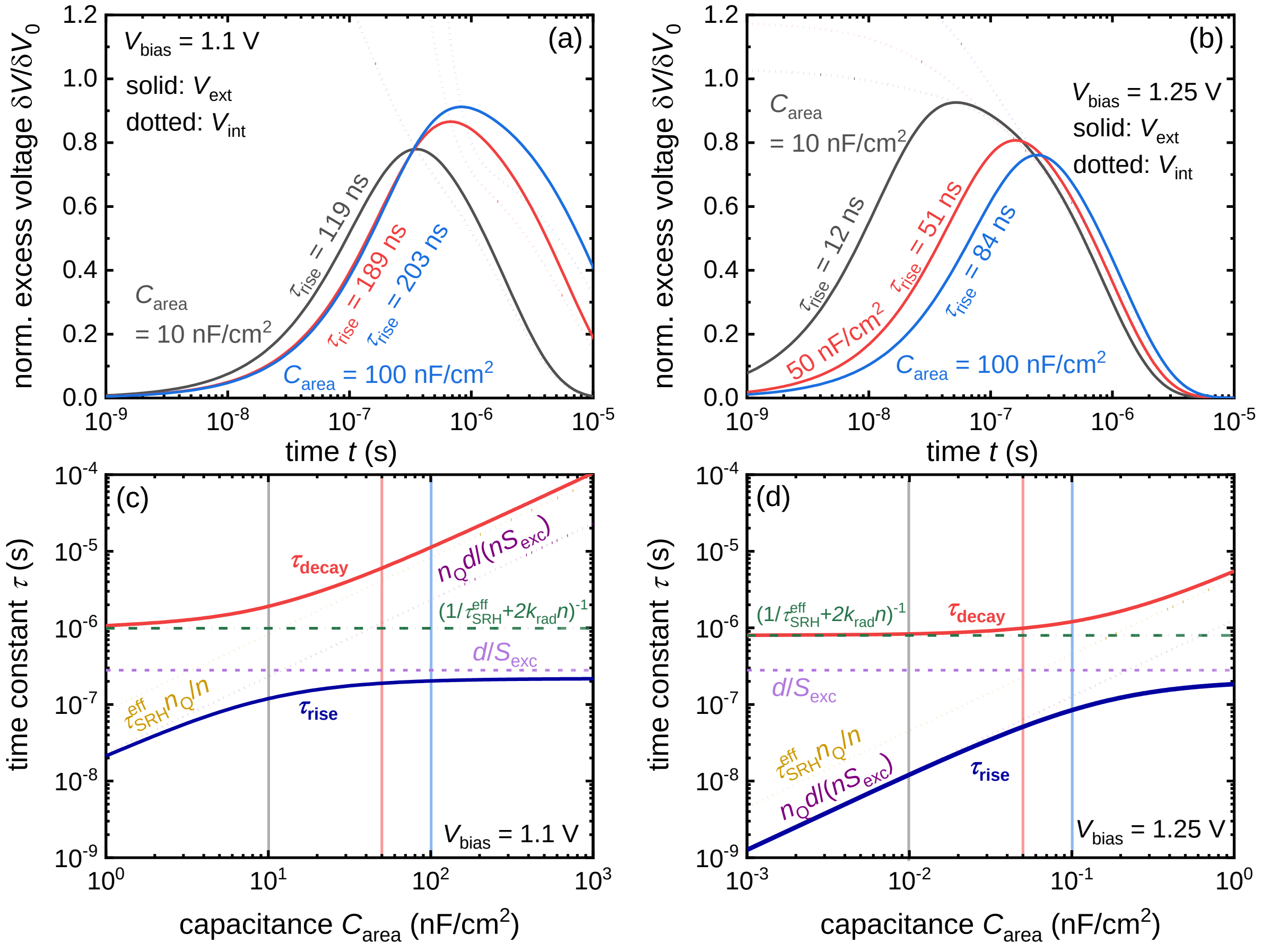

**Figure 3**: (a, b) Analytical solutions of the normalized internal excess-voltage decay $\delta V_{\text{int}}(t)/\delta V_0$ (dotted line) and the respective normalized external excess-voltage curve of $\delta V_{\text{ext}}(t)/\delta V_0$ (solid line) at (a) 1.1 V and (b) 1.25 V bias voltage, calculated for three different values of the electrode capacitance per area $C_{\text{area}}$. The capacitance per area $C_{\text{area}}$ is varied between 20 nF/cm$^2$ (grey), 50 nF/cm$^2$ (red) and 100 nF/cm$^2$ (blue). Besides a constant exchange velocity of $S_{\text{exc}}$ = 100 cm/s, other simulation parameters are equal to Figure 2. (c, d) Rise time $\tau_{\text{rise}}$ and decay time $\tau_{\text{decay}}$ from equation 9 as a function of the electrode capacitance per area $C_{\text{area}}$ for the two different bias levels. Furthermore, additional time constants related to the eigenvalues of the matrix are shown as guide to the eye.

Figure 3(a) and (b) show the normalized internal excess voltage $\delta V_{\text{int}}(t)/\delta V_0$ (dotted line) and the respective external voltage $\delta V_{\text{ext}}(t)/\delta V_0$ (solid line) calculated from equation (8) for two different bias levels. In addition, the lower two bottom panels (c) and (d) again plot the time constants of the rise and decay, but this time as a function of capacitance $C_{area}$. Higher capacitances imply that more charge carriers have to be transferred from the perovskite to the electrodes to accommodate a given change in external voltage. Thus, a higher value of $C_{area}$ slows down both the rise and fall of the normalized internal and external voltage curves, reduces the height of the $\delta V_{ext}(t)/\delta V_0$ peak and also affects the shape of the curves. Moreover, the rise- and decay-time curves again consist of two regions for both shown bias voltages. In one region $\tau_{decay}$ and $\tau_{rise}$ are constant, while they are following a power law and increase for higher $C_{area}$ in the other one. The decay time $\tau_{decay}$ does not depend on $C_{area}$ as long as the time constant of electrode discharging is small in comparison to the recombination lifetime $\tau_R$. In this regime, which almost disappears in case of lower bias voltage of 1.1 V, $\tau_{decay}$ is limited by the green, dashed line. In the second regime, the decay time increases with higher electrode capacitance $C_{area}$ and follows the yellow, dotted line of $\tau_{SRH}^{eff} n_Q / n_{bias}$ . The rise time $\tau_{rise}$ is proportional to $dn_Q / (S_{exc} n_{bias})$ and therefore increases with higher capacitance until this time constant gets larger than the *RC*-element formed by the resistance of charge extraction and the chemical capacitance of the absorber layer. The *RC*-limitation that sets the saturation value of the rise time is not universal and depends on the relations of the respective time constants to each other.

We observed in Figures 2 and 3 that in addition to parameters such as $C_{area}$, $S_{exc}$, $k_{rad}$ and $\tau_{SRH}^{eff}$ also the bias voltage affects the time constants for rise and decay. Thus, Figure 4 shows the rise time $\tau_{rise}$ and decay time $\tau_{decay}$ as a function of the bias open-circuit voltage $V_{oc}$. The three Figures 4(a)-(c) on the left-hand side show the time constants versus $V_{oc}$ for three voltage-independent values of $S_{exc}$, (1000 cm/s, 100 cm/s, 10 cm/s). The decay time $\tau_{decay}$, (red line) shows three distinct regions, the capacitively limited region at low $V_{oc}$, the SRH dominated region at intermediate $V_{oc}$ and the region limited by radiative recombination at high $V_{oc}$.[42] The rise time $\tau_{rise}$ (blue line) shows two distinct regions. At low bias voltages the $\tau_{rise}$ is constant, while at high bias voltages it continuously decreases with $\tau_{rise} \sim 1/n_{bias}$. What changes significantly for the three different exchange velocities is the position and distance of the two time-constant curves relative to each other. Moreover, depending on the exchange velocity, different *RC*-time constants dominate the regimes and saturation values of $\tau_{rise}$ and $\tau_{decay}$.

As the speed of charge injection and extraction via the transport layers depends on the electric field, it is realistic to expect $S_{exc}$ to depend on bias voltage. While there are analytical approximations for $S_{exc}$, those are valid close to short circuit and inaccurate close to open circuit. Thus, we calculate the voltage dependence of $S_{exc}$ from numerical simulations of device current-voltage curves as described in section III.f of the SI. Figures 4(d)-(f) show the results for voltage-dependent values of $S_{exc}$. The panels (d) to (f) differ in the assumed values of the electron and hole mobilities in the transport layers ($\mu = 10^{-2}$ cm$^2$/Vs in (d) to $\mu = 10^{-4}$ cm$^2$/Vs in (f)). While the result for $\tau_{decay}$ does not significantly differ from the case of a voltage independent $S_{exc}$, the voltage dependence of $\tau_{rise}$ changes considerably. The rise time now depends less strongly on bias voltage and shows a relatively flat plateau at intermediate voltages.

Figure 4 shows that rise and decay times are dominated by several different terms depending on the relative values of the different time constants. The fairly high complexity of equations (9) and (10) however does not give direct access to the important time constants. To understand and identify the most important time constants for rise and decay it is therefore useful to simplify equation (9) further.

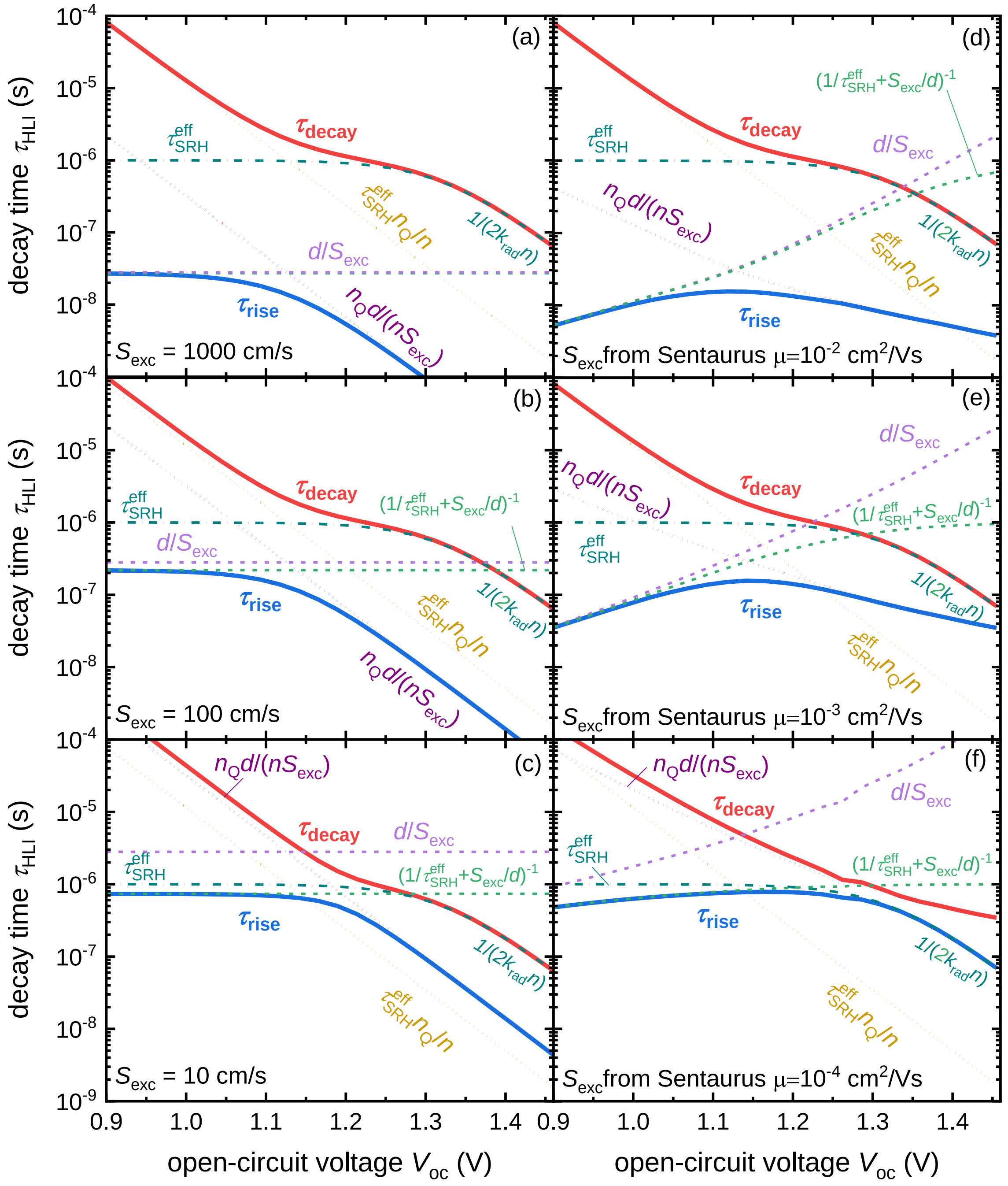

**Figure 4**: Decay time $\tau_{decay}$ (red) and rise time $\tau_{rise}$ (blue) as a function of the open-circuit voltage $V_{oc}$, calculated using equation (9). In addition, the inverse coefficients $1/k_2 = d/S_{exc}$ and $1/k_3 = dn_Q/(S_{exc}n_{bias})$ are plotted, as well as additional important time constants, like $\tau_{SRH}^{eff} n_Q \big/ n_{bias}$ which is related to electrode discharging. The figures in the left column show the analytical results using a constant exchange velocity $S_{exc}$ of (a) $10^3$ cm/s, (b) 100 cm/s and (c) 10 cm/s. In contrast the Figures (d), (e) and (f) on the right-hand side show the decay time $\tau_{decay}$ and rise time $\tau_{rise}$ for three different voltage-dependent exchange velocities $S_{exc}$.

A Taylor series expansion of equation (10) allows us to mathematically approximate the decay time $\tau_{decay}$ and simplify it to the point where the behavior that is described above can be read directly from the equation (for details see section III.d in the supporting information). In this approximation, the decay time results from

$$\tau_{\text{decay}} \approx \frac{k_1+k_2+k_3}{k_1k_3} = \left(\frac{1}{\tau_{\text{R}}}+\frac{S_{\text{exc}}}{d}+\frac{S_{\text{exc}}n_{\text{bias}}}{dn_{\text{Q}}}\right)\Bigg/\left(\frac{1}{\tau_{\text{R}}}\frac{S_{\text{exc}}n_{\text{bias}}}{dn_{\text{Q}}}\right) = \frac{dn_{\text{Q}}}{S_{\text{exc}}n_{\text{bias}}}+\frac{\tau_{\text{R}}n_{\text{Q}}}{n_{\text{bias}}}+\tau_{\text{R}}\ , \quad (14)$$

which corresponds to a series connection of the recombination lifetime and both *RC*-time constants from electrode discharging and charge transfer. As long as the exchange velocity is relatively high ( $\tau_{\text{SRH}}^{\text{eff}} > d/S_{\text{exc}}$ ), equation (14) can be simplified even further, resulting in

$$\tau_{\text{decay}} \approx \tau_{\text{R}}\left(1+\frac{n_{\text{Q}}}{n_{\text{bias}}}\right) = \frac{n_{\text{Q}}/n_{\text{bias}}+1}{2k_{\text{rad}}n_{\text{bias}}+1/\tau_{\text{SRH}}^{\text{eff}}} \quad (15)$$

and thereby matches equation (12) of our previous publication.[42] Using a similar mathematical approach, also allows us to derive an approximation for the rise time

$$\tau_{\text{rise}} \approx \left(k_1+k_2+k_3\right)^{-1} = \left(\frac{1}{\tau_{\text{R}}}+\frac{S_{\text{exc}}}{d}+\frac{S_{\text{exc}}n_{\text{bias}}}{dn_{\text{Q}}}\right)^{-1}. \quad (16)$$

Basically, this rise-time approximation is a parallel connection of the recombination lifetime $\tau_R$, the *RC*-time constant $d/S_{exc}$ given by the resistance of charge transfer multiplied with the chemical capacitance of the perovskite, and $dn_{\text{Q}}/\left(S_{\text{exc}}n_{\text{bias}}\right)$, being the *RC*-time constant of charge-transfer resistance and electrode capacitance. As long as the condition $\tau_R > d/S_{exc}$ holds, equation (16) can be simplified even further, resulting in

$$\tau_{\text{rise}} \approx \frac{d}{S_{\text{exc}}\left(1+\frac{n_{\text{bias}}}{n_{\text{Q}}}\right)}, \quad (17)$$

which allows us to calculate $S_{exc}$ from $\tau_{rise}$ and vice versa. Note that equations (15) and (16) provide relatively simple approximations for the rise and decay time, each of which include $n_Q$ and therefore the influence of the electrode capacitance. The electrode capacitance has a strong impact on transient experiments performed on complete solar cells but has no direct impact on the steady state functionality of the device. Thus, we will now investigate, how we can extract the performance limiting contributions to the rise and decay time from the capacitive contributions.

## V. Application of the Analytical Two-Component Model to Experimental TPV Data

Figure 5(a) shows simulated values of $S_{exc}$ resulting from different numerical drift-diffusion simulations (lines) where the mobility of electrons and holes in the charge transport layers was varied. This data is compared to values of $S_{exc}$ resulting from the application of equation (17) to the experimental data on a MAPI solar cell shown in Figure 5(b). The experimental data shows that the recombination lifetime $\tau_R$ is approximately one order of magnitude higher than $\tau_{rise}$, which implies that equations (15) and (17) are both good approximations for this particular device.

As shown in the supporting information (section III.g) and based on the rationale derived in ref. [49], it is possible to use the ratio of the two time constants $\tau_R$ and $d/S_{exc}$ to estimate the effect of slow charge extraction on the current-voltage curve and thereby performance of a solar cell. The steady state equivalent of our two-component model allows us to calculate the current-voltage curve via

$$J = qd\left(R-G\right) = qd\left(\frac{1}{1+d/\left(S_{exc}\tau_R\right)}\right)\left[\frac{n_0}{\tau_R}\exp\left(\frac{qV_{ext}}{2k_BT}\right)-G\right] \tag{18}$$

If the product $\tau_R S_{exc} >> d$, charge extraction is efficient, and equation (18) predicts that the *JV*-curve only depends on recombination and generation (square bracket) but not on the efficiency of charge extraction (round bracket). If however $\tau_R S_{exc} \approx d$, the term in the round bracket becomes ½ and the current density is reduced by 50%. Thus, a low value of $S_{exc}$ will reduce $J_{sc}$ by the factor $\left(1+d/\left(S_{exc}\tau_R\right)\right)^{-1}$. A voltage dependence of this factor can then also reduce the fill factor. In contrast, as open circuit is not affected by charge extraction, the open-circuit voltage remains independent of $S_{exc}$.

In order to quantify the efficiency of charge extraction, we have two possibilities. Option (1) is to plot the ratio of the two time constants, i.e. $\tau_R S_{exc}/d$, which is a quantity that should be significantly larger than unity to ensure efficient collection. Alternatively, we can directly compute the factor $\left(1+d/\left(S_{exc}\tau_R\right)\right)^{-1}$ which is a quantity lower than one that defines the current loss due to inefficient charge collection. Both figures of merit (FOMs) are shown in Figures 5(c) and (d) using the experimental data for the rise and decay time shown in Figure 5(b). Here, the first figure of merit follows from

$$FOM_1 = \frac{S_{exc}\tau_R}{d} = \frac{\tau_{decay}}{\tau_{rise}}\frac{n_Q n_{bias}}{\left(n_Q + n_{bias}\right)^2} \tag{19}$$

while the second one can be obtained via

$$FOM_2 = \frac{1}{1+\dfrac{d}{S_{exc}\tau_R}} = \frac{1}{1+\dfrac{\tau_{rise}}{\tau_{decay}}\dfrac{\left(n_Q + n_{bias}\right)^2}{n_Q n_{bias}}}. \tag{20}$$

Both FOMs are relatively constant between 0.9 and 1.1 V, where $FOM_1 \approx 10$ implying that extraction is significantly faster than recombination. Towards higher voltages, $FOM_1$ significantly reduces and quickly falls below unity. As can be seen in Figure 5(b) this drop is mostly due to $d/S_{exc}$ getting significantly longer towards higher voltages. We assume that this is due to the fact that at voltages above 1.2 V likely no significant electric field is left in the transport layers. Potentially, there are even extraction barriers implying that electrons and holes would have to diffuse against the electric field to be extracted. Figure 5(e) illustrates the impact of the FOMs on the current-voltage curve. The solid line shows the experimentally measured current voltage curve of the MAPI solar cell whose TPV data is shown in Figure 5(b). The yellow circles show the *JV*-curve according to equation (18) but assuming $S_{exc}$ → ∞. The blue spheres show the current voltage curve according to equation (18) with $FOM_2$ according to Figure 5(d). We note that the *JV* curve shows a slightly reduced $J_{sc}$ and FF. The change in $J_{sc}$ is due to a $FOM_1 \approx 10$ leading to a $FOM_2 \approx 0.9$, implying a still quite significant loss in $J_{sc}$ of just less than 10%. The *FF* is only moderately affected as $FOM_2$ is relatively voltage independent up to about 1.1 V. The real *JV*-curve obviously suffers from additional losses not included in $FOM_2$. We assume that these additional losses are ohmic series resistance losses, e.g. due to lateral transport in the transparent conductive oxide. As ohmic losses due to lateral transport would not be included in $S_{exc}$, it is plausible that the blue spheres do not reproduce the *JV*-curve. We added an ohmic series resistance ($R_s = 4\ \Omega\text{cm}^2$) to this curve and obtain the open circles, which approximate the experimental *JV*-curve well.

Thus, we conclude that the FOMs are consistent with the *JV*-curve but would require additional information to predict the *JV*-curve. We also show that due to the rather low voltage dependence of $S_{exc}$ up to the maximum power point, extraction losses are mostly affecting the $J_{sc}$ and to a lesser degree the *FF*. This also implies that the *FF* is most likely more affected by ohmic series resistances than by the non-ohmic charge transport through ETL and HTL. This observation is consistent with other experimental findings on charge extraction such as the weak voltage dependence of steady-state photoluminescence between short circuit and the maximum-power point in perovskite solar cells.[16, 45]

## VI. Conclusions

A significant shortcoming of previous approaches to analyse small signal transient photovoltage data was that they were generally focussed on obtaining only the decay time constant. While it was possible to determine rise times, there was no model to relate the rise times to device functionality. Here, we introduce a model for charge recombination and extraction that can be linearized and solved analytically. The solutions for the rise and decay time of the transient photovoltage follow from the inverse eigenvalues of a matrix and can be related to physical mechanisms such as extraction, recombination and capacitive charging and discharging of the electrodes. We apply the model to experimental data and identify the physical mechanisms that determine the rise and decay times at different bias conditions. From the absolute value, their bias dependence and ratio of the rise and decay times, we can clearly distinguish the time constants of recombination and extraction from time constants related to capacitive effects that are irrelevant for steady state device performance. By correcting the rise and decay times for the influence of the electrode capacitance, we obtain figures of merit for charge extraction that connect the data obtained from the transient with the current-voltage curves and thereby the performance of the device. We observe that the rise and decay times predict losses in $J_{sc}$ of about 10% as well as minor losses in *FF*.

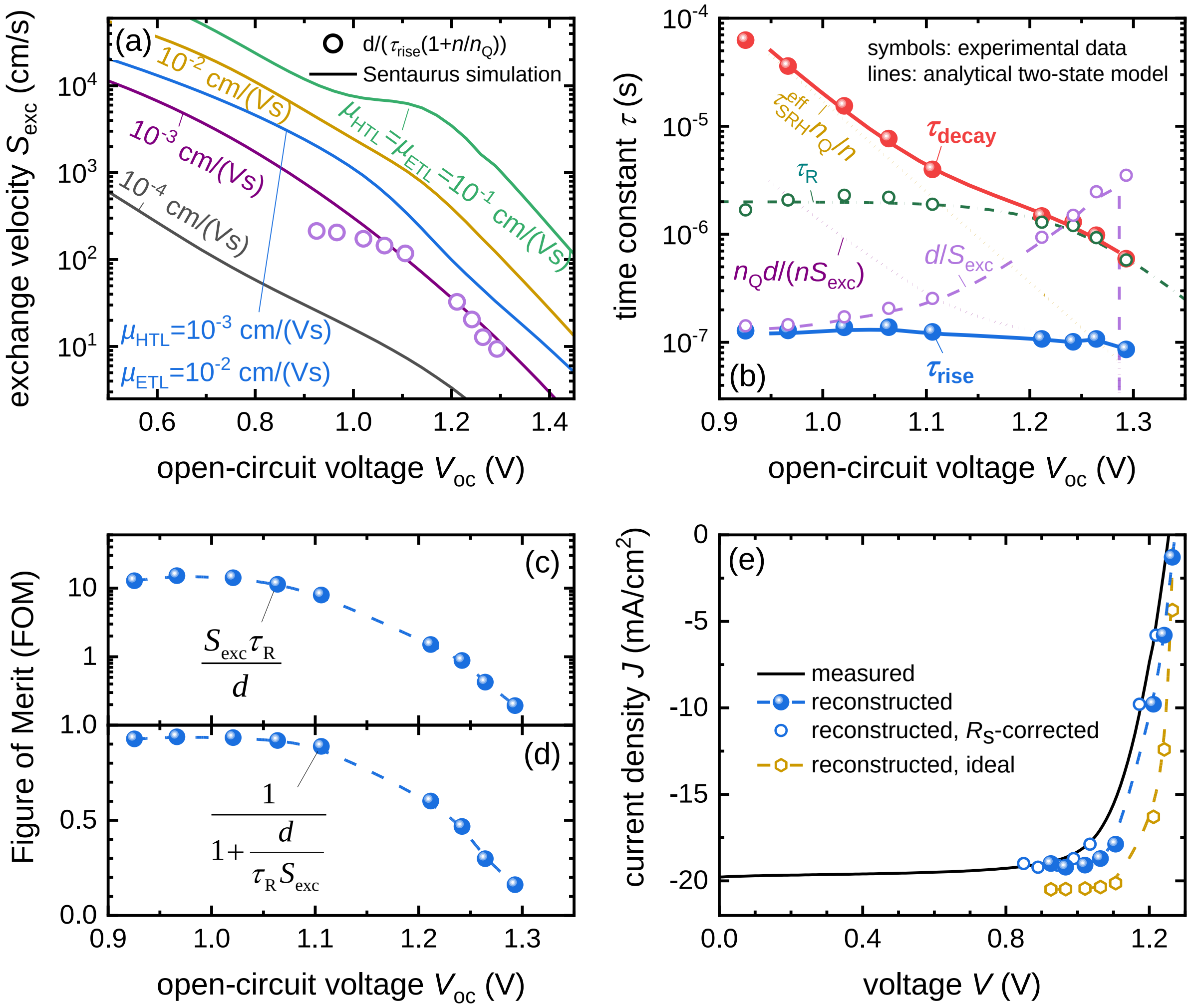

**Figure 5**: (a) Exchange velocity $S_{exc}$ extracted from steady state drift-diffusion simulations of *JV*-curves, being conducted assuming various charge-carrier mobilities. Moreover, the exchange velocity $S_{exc}$ calculated from the experimental TPV data via $S_{exc} = d/\left(\tau_{rise}\left(1+n_{bias}/n_Q\right)\right)$ is shown by the dotted symbols. (b) Experimental data of the decay time $\tau_{decay}$ (red symbols) and rise time $\tau_{rise}$ (blue symbols) from TPV as a function of the open-circuit voltage $V_{oc}$. In addition, the analytical two-component model is fitted to experimental data. The results are shown by the lines. (c,d) Figures of merit for the rate of extraction vs. recombination. Panel (c) shows the ratio of the time constants. Panel (d) shows the prefactor in equation (18) that quantifies the effect of a finite efficiency of extraction on the *JV* curve. (e) Measured and reconstructed current-voltage characteristic of the MAPI solar cell. In (e), we assumed a generation rate $G = 4.57\times10^{21}$ $cm^{-3}s^{-1}$ and a series resistance $R_s = 4\ \Omega cm^2$.

**Author Information**

Corresponding Author:

*To whom correspondence should be addressed. E-Mail: t.kirchartz@fz-juelich.de, u.rau@fz-juelich.de

**Acknowledgements**

The authors acknowledge support from the Helmholtz Association via the project PEROSEED and via the project-oriented funding (POF IV). We also acknowledge funding from the DFG for the project CREATIVE within the SPP "Perovskite Semiconductors: From Fundamental Properties to Devices" (SPP 2196). Open access publication funded by the German Research Foundation (DFG) – 491111487.

**References**

[1] J. Jeong, M. Kim, J. Seo, H. Lu, P. Ahlawat, A. Mishra, Y. Yang, M. A. Hope, F. T. Eickemeyer, M. Kim, Y. J. Yoon, I. W. Choi, B. P. Darwich, S. J. Choi, Y. Jo, J. H. Lee, B. Walker, S. M. Zakeeruddin, L. Emsley, U. Rothlisberger, A. Hagfeldt, D. S. Kim, M. Grätzel, J. Y. Kim, *Nature* **2021,** *592*, 381.

[2] J. J. Yoo, G. Seo, M. R. Chua, T. G. Park, Y. Lu, F. Rotermund, Y.-K. Kim, C. S. Moon, N. J. Jeon, J.-P. Correa-Baena, V. Bulović, S. S. Shin, M. G. Bawendi, J. Seo, *Nature* **2021,** *590*, 587.

[3] O. Almora, D. Baran, G. C. Bazan, C. I. Cabrera, S. Erten-Ela, K. Forberich, F. Guo, J. Hauch, A. W. Y. Ho-Baillie, T. J. Jacobsson, R. A. J. Janssen, T. Kirchartz, N. Kopidakis, M. A. Loi, R. R. Lunt, X. Mathew, M. D. McGehee, J. Min, D. B. Mitzi, M. K. Nazeeruddin, J. Nelson, A. F. Nogueira, U. W. Paetzold, B. P. Rand, U. Rau, H. J. Snaith, E. Unger, L. Vaillant-Roca, C. Yang, H.-L. Yip, C. J. Brabec, *Advanced Energy Materials* **2023,** *n/a*, 2203313.

[4] M. Kim, J. Jeong, H. Lu, T. K. Lee, F. T. Eickemeyer, Y. Liu, I. W. Choi, S. J. Choi, Y. Jo, H.-B. Kim, S.-I. Mo, Y.-K. Kim, H. Lee, N. G. An, S. Cho, W. R. Tress, S. M. Zakeeruddin, A. Hagfeldt, J. Y. Kim, M. Grätzel, D. S. Kim, *Science* **2022,** *375*, 302.

[5] Q. Jiang, J. Tong, Y. Xian, R. A. Kerner, S. P. Dunfield, C. Xiao, R. A. Scheidt, D. Kuciauskas, X. Wang, M. P. Hautzinger, R. Tirawat, M. C. Beard, D. P. Fenning, J. J. Berry, B. W. Larson, Y. Yan, K. Zhu, *Nature* **2022,** *611*, 278.

[6] Z. Li, B. Li, X. Wu, S. A. Sheppard, S. Zhang, D. Gao, N. J. Long, Z. Zhu, *Science* **2022,** *376*, 416.

[7] K. Yoshikawa, H. Kawasaki, W. Yoshida, T. Irie, K. Konishi, K. Nakano, T. Uto, D. Adachi, M. Kanematsu, H. Uzu, K. Yamamoto, *Nature Energy* **2017,** *2*, 17032.

[8] K. Yoshikawa, W. Yoshida, T. Irie, H. Kawasaki, K. Konishi, H. Ishibashi, T. Asatani, D. Adachi, M. Kanematsu, H. Uzu, K. Yamamoto, *Solar Energy Materials and Solar Cells* **2017,** *173*, 37.

[9] C. Ballif, F.-J. Haug, M. Boccard, P. J. Verlinden, G. Hahn, *Nature Reviews Materials* **2022**.

[10] Y. Rong, Y. Hu, A. Mei, H. Tan, M. I. Saidaminov, S. I. Seok, M. D. McGehee, E. H. Sargent, H. Han, *Science* **2018,** *361*, eaat8235.

[11] D. Zhang, D. Li, Y. Hu, A. Mei, H. Han, *Communications Materials* **2022,** *3*, 58.

[12] M. Stolterfoht, M. Grischek, P. Caprioglio, C. M. Wolff, E. Gutierrez-Partida, F. Peña-Camargo, D. Rothhardt, S. Zhang, M. Raoufi, J. Wolansky, M. Abdi-Jalebi, S. D. Stranks, S. Albrecht, T. Kirchartz, D. Neher, *Advanced Materials* **2020,** *32*, 2000080.

[13] L. Krückemeier, U. Rau, M. Stolterfoht, T. Kirchartz, *Advanced Energy Materials* **2020,** *10*, 1902573.

[14] S. Cacovich, G. Vidon, M. Degani, M. Legrand, L. Gouda, J.-B. Puel, Y. Vaynzof, J.-F. Guillemoles, D. Ory, G. Grancini, *Nature Communications* **2022,** *13*, 2868.

[15] M. Stolterfoht, P. Caprioglio, C. M. Wolff, J. A. Marquez, J. Nordmann, S. Zhang, D. Rothhardt, U. Hörmann, Y. Amir, A. Redinger, L. Kegelmann, F. Zu, S. Albrecht, N. Koch, T. Kirchartz, M. Saliba, T. Unold, D. Neher, *Energy & Environmental Science* **2019,** *12*, 2778.

[16] M. Stolterfoht, V. M. Le Corre, M. Feuerstein, P. Caprioglio, L. J. A. Koster, D. Neher, *ACS Energy Letters* **2019,** 2887.

[17] W. Tress, M. Yavari, K. Domanski, P. Yadav, B. Niesen, J. P. Correa Baena, A. Hagfeldt, M. Graetzel, *Energy & Environmental Science* **2018,** *11*, 151.

[18] P. Caprioglio, C. M. Wolff, O. J. Sandberg, A. Armin, B. Rech, S. Albrecht, D. Neher, M. Stolterfoht, *Advanced Energy Materials* **2020,** *10*, 2000502.

[19] L. Krückemeier, B. Krogmeier, Z. Liu, U. Rau, T. Kirchartz, *Advanced Energy Materials* **2021,** *11*, 2003489.

[20] E. M. Hutter, T. Kirchartz, B. Ehrler, D. Cahen, E. v. Hauff, *Applied Physics Letters* **2020,** *116*, 100501.

[21] T. Kirchartz, J. A. Márquez, M. Stolterfoht, T. Unold, *Advanced Energy Materials* **2020,** *10*, 1904134.

[22] F. Staub, H. Hempel, J. C. Hebig, J. Mock, U. W. Paetzold, U. Rau, T. Unold, T. Kirchartz, *Physical Review Applied* **2016,** *6*, 044017.

[23] S. Feldmann, S. Macpherson, S. P. Senanayak, M. Abdi-Jalebi, J. P. H. Rivett, G. Nan, G. D. Tainter, T. A. S. Doherty, K. Frohna, E. Ringe, R. H. Friend, H. Sirringhaus, M. Saliba, D. Beljonne, S. D. Stranks, F. Deschler, *Nature Photonics* **2020,** *14*, 123

[24] J. M. Richter, M. Abdi-Jalebi, A. Sadhanala, M. Tabachnyk, J. P. H. Rivett, L. M. Pazos-Outon, K. C. Gödel, M. Price, F. Deschler, R. H. Friend, *Nature Communications* **2016,** *7*, 13941.
[25] D. Kiermasch, L. Gil-Escrig, A. Baumann, H. J. Bolink, V. Dyakonov, K. Tvingstedt, *Journal of Materials Chemistry A* **2019,** *7*, 14712.
[26] D. Kiermasch, A. Baumann, M. Fischer, V. Dyakonov, K. Tvingstedt, *Energy & Environmental Science* **2018,** *11*, 629.
[27] Z. S. Wang, F. Ebadi, B. Carlsen, W. C. H. Choy, W. Tress, *Small Methods* **2020,** *4*, 2000290.
[28] A. O. Alvarez, S. Ravishankar, F. Fabregat-Santiago, *Small Methods* **2021,** *n/a*, 2100661.
[29] S. Ravishankar, A. Riquelme, S. K. Sarkar, M. Garcia-Batlle, G. Garcia-Belmonte, J. Bisquert, *The Journal of Physical Chemistry C* **2019,** *123*, 24995.
[30] J. Bisquert, M. Janssen, *The Journal of Physical Chemistry Letters* **2021,** *12*, 7964.
[31] J. Bisquert, *The Journal of Physical Chemistry Letters* **2022,** *13*, 7320.
[32] A. Guerrero, J. Bisquert, G. Garcia-Belmonte, *Chemical Reviews* **2021,** *121*, 14430.
[33] A. Kiligaridis, P. A. Frantsuzov, A. Yangui, S. Seth, J. Li, Q. An, Y. Vaynzof, I. G. Scheblykin, *Nature Communications* **2021,** *12*, 3329.
[34] F. Peña-Camargo, J. Thiesbrummel, H. Hempel, A. Musiienko, V. M. L. Corre, J. Diekmann, J. Warby, T. Unold, F. Lang, D. Neher, M. Stolterfoht, *Applied Physics Reviews* **2022,** *9*, 021409.
[35] B. Krogmeier, F. Staub, D. Grabowski, U. Rau, T. Kirchartz, *Sustainable Energy & Fuels* **2018,** *2*, 1027.
[36] J. Haddad, B. Krogmeier, B. Klingebiel, L. Krückemeier, S. Melhem, Z. Liu, J. Hüpkes, S. Mathur, T. Kirchartz, *Advanced Materials Interfaces* **2020,** *7*, 2000366.
[37] C. Fai, A. J. C. Ladd, C. J. Hages, *Joule* **2022,** *6*, 2585.
[38] M. Azzouzi, P. Calado, A. M. Telford, F. Eisner, X. Hou, T. Kirchartz, P. R. F. Barnes, J. Nelson, *Solar RRL* **2020,** *4*, 1900581.
[39] C. Cho, S. Feldmann, K. M. Yeom, Y.-W. Jang, S. Kahmann, J.-Y. Huang, T. C. J. Yang, *Nature Materials* **2022**.
[40] G. Vidon, S. Cacovich, M. Legrand, A. Yaiche, D. Ory, D. Suchet, J.-B. Puel, J.-F. Guillemoles, *Physical Review Applied* **2021,** *16*, 044058.
[41] S. Wheeler, D. Bryant, J. Troughton, T. Kirchartz, T. Watson, J. Nelson, J. R. Durrant, *The Journal of Physical Chemistry C* **2017,** *121*, 13496.
[42] L. Krückemeier, Z. Liu, B. Krogmeier, U. Rau, T. Kirchartz, *Advanced Energy Materials* **2021,** *11*, 2102290.
[43] P. R. F. Barnes, K. Miettunen, X. Li, A. Y. Anderson, T. Bessho, M. Gratzel, B. C. O'Regan, *Adv. Mater* **2012,** *25*, 1881.
[44] Z. Liu, L. Krückemeier, B. Krogmeier, B. Klingebiel, J. A. Marquez, S. Levcenko, S. Öz, S. Mathur, U. Rau, T. Unold, T. Kirchartz, *ACS Energy Letters* **2019,** *4*, 110.
[45] D. Grabowski, Z. Liu, G. Schöpe, U. Rau, T. Kirchartz, *Solar RRL* **2022,** *6*, 2200507.
[46] U. Rau, *IEEE Journal of Photovoltaics* **2012,** *2*, 169.
[47] O. Breitenstein, *IEEE Journal of Photovoltaics* **2014,** *4*, 899.
[48] J. Bisquert, *Physical Chemistry Chemical Physics* **2003,** *5*, 5360.
[49] U. Rau, V. Huhn, B. E. Pieters, *Physical Review Applied* **2020,** *14*, 014046.